\documentclass[]{aa}  
\usepackage{graphicx}
\usepackage{eqnarray,amsmath}
\usepackage{multicol}
\usepackage{txfonts}
\usepackage{hyperref}
\usepackage{xcolor}
\newcommand{\pp}{\partial}

\newcommand{\laa}{\lambda}

\newcommand{\gaa}{\gamma}

\begin{document} 
	\title{Revisiting Viscous Transonic Decretion Disks of Be Stars}

	\author{Michel Curé\inst{1}
		\and
		Rodrigo Meneses\inst{2}
		\and
		Ignacio Araya\inst{3}
		\and
        Catalina Arcos\inst{1}
		\and
		Greco Pe\~na\inst{1}
		\and
		Natalia Machuca\inst{1}
		\and
		Abigali Rodriguez\inst{1}
}
	%\fnmsep\thanks{Just to show the usage of the elements in the author field}
	
	\institute{Instituto de Física y Astronomía, Universidad de Valparaíso. Av. Gran Breta\~na 1111, Casilla 5030, Valpara\'iso, Chile.\\
		\email{michel.cure@uv.cl}
		\and
		Escuela de Ingenier\'ia Civil, Universidad de Valparaíso. General Cruz 222, Valpara\'iso, Chile.
		\and
		Vicerrectoría de Investigación, Universidad Mayor, Manuel Montt 367, Santiago, Chile}
	
	\date{Received March 2022; accepted 2022}
	
	% \abstract{}{}{}{}{} 
	% 5 {} token are mandatory
	
	\abstract % 5 {} token are mandatory
	{}
	{In the context of Be stars, we re-studied the viscous transonic decretion disk model of these stars. This model is driven by a radiative force due to an ensemble of optically-thin lines and viscosity considering the Shakura–Sunyaev prescription.}
	%{We re-studied the viscous transonic decretion disks of Be stars, driven by viscosity and a radiative force due to an ensemble of optically-thin lines with the Shakura–Sunyaev viscosity prescription. }
	{The non-linear equation of motion  presents a singularity (sonic point) and an eigenvalue, which is also the initial condition at the stellar surface. Then, to obtain this eigenvalue, we set it as a radial quantity and perform a detailed topological analysis. Thereafter, we describe a numerical method for solving either Nodal and Saddle transonic solutions.}
	{The value of the viscosity, $\alpha$, barely determine the location of the sonic point, but it determines the topology of the solution. We found two Nodal solutions, which are almost indistinguishable between them. Saddle solutions are founded for lower values of $\alpha$ than the required of the Nodal solutions.
	In addition, rotational velocity do not play a determine role in the velocity (and density) profile, because viscosity effects collapse all the solutions to almost a unique one in a small region above the stellar surface.}
	{A suitable combination of line-force parameters and/or disk temperature, give location of the sonic point lower than 50 stellar radii, describing a truncated disk. This could explain the SED turndown observed in Be stars without needing a binary companion.}
	
	\maketitle
	
%_____INTRODUCTION________________________________________________________________________________
%\begin{multicols}{1}
\section{Introduction}
Classical Be stars (CBes) are fast rotating main sequence B-type stars forming an equatorial gas rotating disk. The inner part of the disk, usually $\lesssim$ 20 stellar radii, is geometrically thin and optically thick and rotates in a quasi-Keplerian orbit \citep{Quirrenbach1997,Meilland2012,Rivinius2013}. These disks, referred as decretion disks, are built from mass ejected from the equatorial stellar surface that acquires sufficient velocity and angular momentum to orbit the star. Once the material is ejected from the star, is generally governed by gravity and viscous forces \citep[for the latest review see][and references therein]{Rivinius2013}. 

Currently, the best theoretical framework describing the evolution of these disks once formed is the Viscous Decretion Disk model (VDD) proposed by \cite{lee1991}. In this model, viscosity acts to shuffle the angular momentum of the circumstellar material and with the assumption of the gas motion as Keplerian, a steady and thermally-stable structure is obtained. The strength of viscosity is parameterized as $\alpha$, according to the $\alpha$-disk model proposed by \cite{Shakura1973} for accretion disks. In the case of decretion disks, the mass loss rate in charge of feeding the disk has an opposite sign. The $\alpha$ parameter dictates the timescales over which Be star decretion disks evolve and dissipate. In the work made by \cite{Haubois2012} they used VDD model to study the temporal evolution of the disk density in Be stars and found a small fraction ($\sim 1\%$) of the mass ejected by the star acquires sufficient angular momentum to move outwards, orbiting at increasing radii, while a majority of the ejected mass falls back onto the star. \cite{Rimulo2018} analyzed 81 outburst events in 54 CBes by modeling light curves with models that solve the radiative transfer problem in the non-local thermodynamic equilibrium regime in 2 or 3 dimensions. They found the viscosity parameter being larger during disk build-up ($\alpha$=0.63) in comparison with the  dissipation phase ($\alpha$=0.29). This means that the formation timescale is faster than the dissipation phase. 
In addition to viscosity, radiative ablation may systematically remove, by means of radiative acceleration, some material from the inner part of CBes disks, thus forming a disk-wind type of structure both above and below the disk \citep[][]{Kee2016,Kee2018a,Kee2018b,Kee2019}. 

On the other hand, \cite{okazaki2001} solved the hydrodynamical equations for a viscous disk through the $\alpha$-disk model, assuming isothermal conditions and a radiative force given by an ensemble of optically-thin lines. His goal was to find the transition between near-Keplerian and angular momentum conserving motion inside the disk. Since the specific angular momentum increases with the distance from the stellar photosphere making it difficult to hold a Keplerian motion at larger distances. He found for three different $\alpha$ values (1.0, 0.1, and 0.01) the existence of a transonic solution. The sonic point was located far from the stellar surface at values larger than 100 stellar radii. Inside this region, the outflow was found to be highly subsonic. For a particular set of line-force parameters \cite{okazaki2001} found that for $\alpha \geq 0.95$ the topology of the sonic point is Nodal and for $\alpha < 0.9$ is Saddle. Finally, he found that in the inner subsonic region the disk is near-Keplerian, while in the outer subsonic and supersonic regions the angular momentum is conserved. We note that in his work the specific angular momentum at the sonic point $\ell_{s}$ is set as an eigenvalue, making the angular velocity a fixed quantity. 

\cite{KOM2011} studied the mass-loss rate occurring inside the decretion disks which stems from the angular momentum loss due to the condition of maintain the star rotating close and below their critical rotational speed. Then, from a specific angular momentum, the mass-loss rate is related with the outer disk size. In their work, they discussed three different physical processes that affect the outer disk: binarity, thermal expansion in supersonic flows and radiative ablation in the inner disk, and also, they presented how to implement these considerations about the mass-loss rates in decretion disks in stellar evolution codes. Other study about the outer disk in Be stars was the work from \cite{Kurfurst2014}, where they studied the dependency between the physical features of large disks and their temperature and viscosity distribution. In spite of the relevance of the results of these both mentioned works, in their calculations they do not include radiation force in the wind equations.

\cite{Klement17} noted the importance to extend the study of the density structure using VDD in the outer parts of the disk. Since the majority of observational constraints come from optical and IR wavelengths, the outer regions are detectable only at radio wavelengths. In their work, they compiled data from ultraviolet up to radio wavelengths for the spectral energy distribution and found a turndown in the SED at 10-60$\mu m$ in a sample of 6 Be stars. To reproduce the observations considering a VDD model, they used a truncated disk with radii between 26 to 108 stellar radii. They concluded that tidal forces from a binary companion are the only mechanism (at close distances to the star) that can truncate the disk.

In this paper, we revisited the work of  \cite{okazaki2001} but considered another solution schema that does not involve any angular variable as an eigenvalue, and which will finally lead us to conclusions that can respond to more recent works about observational characteristics in the flux distribution of Be stars, as well as the effects of rotational speed and in the velocity and density wind structure. 

This work is organized as follows. Section 2 reviews the VVD model and propose a new procedure to solve the equation of motion. In section 3 we analyze the topology at the singularity from the equation of motion. Then, in section 4, a numerical procedure is proposed to obtain the different viscous transonic decretion disk solutions. In section 5 numerical calculations are performed to understand the influence of the different parameters. Finally, in sections 6 and 7 we give a discussion and our conclusion, respectively.

\section{Viscous Decretion Disks Hydrodynamic Model}
Based on the VDD scenario proposed by \cite{lee1991},  \cite{okazaki2001} studied the influence of a radiative force due to an ensemble of optically-thin lines described by the ad-hoc model from \cite{chen1994}. In addition, the Shakura–Sunyaev’s prescription for the viscous stress was adopted.\\
For the sake of completeness, we will follow the same nomenclature and equations  from \citet[][see derivation details therein]{okazaki2001}. The geometrically thin circumstellar disk of a Be star is in steady state, is symmetric about the rotational axis and the equatorial plane. 

Then, the disk equations in cylindrical coordinates ($r$, $\phi$, $z$) follow mass conservation: 
\begin{equation}
\dot{M}+2 \pi \,r \,V_r\, \Sigma =0 , 
\label{eq1}
\end{equation}
where $\dot{M}$ is the mass loss  rate, $\Sigma$ is the vertical  integrated density and $V_r$ is the radial component of the vertical averaged velocity.\\
\noindent The momentum conservation, radial and angular components are, respectively:
\begin{eqnarray}
- V_r\, \frac{d V_r}{d r} + \frac{V_\phi^2}{r} - \frac{G\,M}{r^2} -  \frac{1}{\Sigma}\frac{d\,W}{d r} + g_{\rm rad} + \frac{3 W}{2 r \Sigma} \, & =& 0\, ,\label{eq2}\\
V_r\, \frac{d V_\phi}{d r} + \frac{V_r\,V_\phi}{r} - \frac{1}{r^2 \, \Sigma}\frac{d\,(r^2\, t_{r \phi})}{d r}  & =& 0\, ,\label{eq3}
\end{eqnarray}
here $V_\phi$ is the angular component of the vertical ($z$-axis) averaged velocity, $M$ is the stellar mass and $G$ the gravitational constant. $W$ is the pressure and $t_{r \phi}$ is the $r-\phi$ component of the viscous stress. The radiative force (vertically averaged) is given by $F_{\rm rad}$ (see below).
The state equation of an ideal gas reads:
\begin{equation}
W\,= c_s^2 \,\Sigma \, ,
\label{eq4}
\end{equation}
where $c_s^2$ is the isothermal sound speed.\\

Finally, the  Shakura–Sunyaev viscosity prescription is given in terms of the $r-\phi$ component of the viscous stress:
\begin{equation}
t_{r \phi} =- \alpha \, W \, ,
\label{eq5}
\end{equation}
where $\alpha$ is the viscosity parameter.
\subsection{The radiative acceleration}
The radiative acceleration description used in the VDD model, was the proposed by \cite{chen1994}, i.e., the radiative acceleration is produced by an ensemble of optically thin lines, namely:
\begin{equation}
g_{\rm rad} = \frac{G M\, \Gamma}{r^2} + 
\frac{G M\,(1-\Gamma)}{r^2} \, \eta \left(\frac{r}{R}\right)^\epsilon,
\label{cheneq}
\end{equation}
where $\Gamma$ is the Eddington factor due to electron scattering. The parameters $\epsilon$ and $\eta$ characterize the  decay rate and magnitude of the radiative (line) force, respectively, and $R$ is the stellar radius.%\textbf{cuales son los valores típicos de epsilon y eta?}

\subsection{Okazaki's solution procedure}
Rearranging equations (\ref{eq1}) - (\ref{eq5}), a constant of motion is found, namely:
\begin{equation}
\ell +\alpha \,c_s^2 \,\frac{r}{V_r} = C\, ,
\label{CoF}
\end{equation}
here $\ell$ is the angular momentum per mass unit (specific angular momentum) and $C$ is a (unknown) constant of motion or eigenvalue.

The solution of this problem relays in propose a suitable form to calculate the constant $C$. \cite{okazaki2001}  adopted the evaluation of $C$ at the sonic point ($r=r_s$), i.e., 
\begin{equation}
C=\ell_s + \alpha \,c_s^2 \,\frac{r_s}{c_s} \, ,
%= \ell +\alpha \,c_s^2 \,\frac{r}{V_r} \, ,
\label{eqOka}
\end{equation}
where $\ell_s$ is the specific angular momentum at the sonic point. Now the eigenvalue of this problem is $\ell_s$.\\
Then, solving from Eq. (\ref{CoF}) for $\ell$ and eliminating $t_{r \phi}$, $W$ and $\Sigma$ from equations (\ref{eq1})–(\ref{eq3}) we obtain Okazaki's equation of motion (hereafter OEoM):
\begin{equation}
\left( 1-\frac{c_s^2}{V_r^2}\right)\, V_r \frac{d V_r}{dr} = \frac{\ell^2}{r^3} + \frac{5}{2} \frac{c_s^2}{r} + g_{\rm eff} \, ,
\label{eom}
\end{equation}
where $\ell$ is given by:
\begin{equation}
\ell= \ell_s + \alpha \, c_s^2\, \left(\frac{r_s}{c_s} -\frac{r}{V_r} \right) \, , 
\label{ls}
\end{equation}
and the effective gravity, $g_{\rm eff}$, reads:
\begin{eqnarray}
g_{\rm eff}  &=& -\frac{G M}{r^2} + g_{\rm rad} \nonumber\\
           &=& -\frac{G M \,(1-\Gamma)}{r^2} \left(1 - \eta \left(\frac{r}{R}\right)^\epsilon \,\right) \, .
\label{geff}
\end{eqnarray}
His work was advocated to solve the set of equations (\ref{eom}) - (\ref{geff}).  

Our main criticism to this approach, is the evaluation of $C$ at the sonic point for $\ell_s$. This means that now $\ell_s$ is the eigenvalue of the OEoM and corresponds to a rotational quantity, i.e., {\it{it fixes}} the value of the stellar rotational velocity. Thus for an individual star with a set of stellar and line-force parameters, we obtain a solution of the OEoM  {\it{only}} for one specific value for the stellar rotational speed.

At the stellar surface, $r=R$, the azimuthal component of the vertically averaged velocity is equal to the stellar rotational velocity, $V_{\phi}(r=R)=V_{\rm rot}\equiv \Omega V_{\rm crit}$.
Here $V_{\rm crit}$ is the critical rotational speed:
\begin{equation}
V_{\rm crit}=\sqrt{\frac{G M (1- \Gamma)}{R}}\, .
\label{vcrit}
\end{equation}

\noindent Thus, the variable $V_\phi /\sqrt{(G M /R)}$ at the stellar surface is %$\Omega V_{\rm crit} /\sqrt{(G M /R)}$, then
\begin{equation}
\frac{V_\phi(R)}{\sqrt{(G M /R)}} = \Omega \sqrt{(1-\Gamma)}.
\label{vphi1}
\end{equation}
For classical Be stars, the Eddington factor $\Gamma$ is a very small value and can be neglected, obtaining
\begin{equation}
\frac{V_\phi(R)}{\sqrt{(G M /R)}} \simeq  \Omega .
\label{vphi2}
\end{equation}

A close inspection of Figure (2b) from \cite{okazaki2001}, specifically at the upper left corner for the variable $V_{\phi}(R) /\sqrt{(G M /R)}$, shows that for different values of the line-force parameters, the rotational speed of the star is also different. In other words, the eigenvalue of the problem, $\ell_s$, that is directly related with the stellar rotational quantity $\Omega$, depends on the line-force parameters  $\eta$ and $\epsilon$.

This solution scheme is completely different to the one used in m-CAK theory  \citep[see e.g,][and references therein]{michel04}, where $\Omega$ is one of the stellar parameters and it is not determined from the solution of the OEoM (Eq. \ref{eom}).

\subsection{A new solution procedure}
In this section, we propose  an alternative procedure to solve the equation of motion, where the stellar rotational speed is an input stellar parameter and not an eigenvalue. In addition, in Okazakis's procedure, the constant $C$ has to be evaluated at the sonic point, at this point the value of the radial velocity is known, but not its location, therefore there are two unknown quantities: $\ell_s$ and $r_s$ that determine the value of $C$.

\subsubsection{Evaluation of the constant of motion $C$}
Then, evaluating $C$ (Eq. \ref{CoF}) at the stellar surface 
$r=R$, we get
\begin{equation}
\label{constantesup}
C=\ell(R)+\alpha c_{s}^{2}\dfrac{R}{V_r(R)},
\end{equation}
where $\ell(R) = \ell_R = R\, V_{\phi} = R \, \Omega \, V_{\rm crit}$ is a known quantity. \\
Thus, $\ell$ has now the following expression:
\begin{equation}\label{constantC}
\ell=\ell_R+\alpha \, c_{s}^{2}\left( \dfrac{R}{V_R}-\dfrac{r}{V_r} \right).
\end{equation}
In this new solution schema, $V_R = V_r(R)$, is the eigenvalue. 

In order obtain a solution of this problem, we have to solve Eq.~\eqref{eom}, together with Eq.~\eqref{geff} and Eq.~\eqref{constantC}. Here we do not have a \textit{typical} eigenvalue problem for the unknown $V_R$, as in the m-CAK theory. In this case, $V_R$ is {\it also} the initial condition at the stellar surface of this non-linear fist-order differential equation.

\subsubsection{Dimensionless variables}
This problem can be expressed in a dimensionless form. First, defining\footnote{We will use indistinctly $u$ or $r$ throughout this work}
$u=-R/r$ and $w(u)=V_r(r)/c_s$ with the following constants
 \begin{equation}
 \label{para}
 \gaa_{1}=\dfrac{GM}{c_{s}^{2}R}, 
 \end{equation}
 \begin{equation} 
 \gaa_{2}=\dfrac{1}{c_{s}^{2}}\left( \dfrac{\ell_R}{R} \right)^{2},
  \end{equation}
 and
 \begin{equation} 
 \gaa_{3}=\dfrac{\alpha c_{s}R}{\ell_R}.
 \end{equation}
Then, substituting Eq.~\eqref{geff} and Eq.~\eqref{constantC} in Eq.~\eqref{eom}, we obtain the dimensionless equation of motion (hereafter EoM):
\begin{equation}
\label{eom2}
%\left\lbrace
%\begin{array}{rcl}
\left(1-w^{2}\right)\dfrac{dw}{du} =w\,F(u,w;\laa) \qquad -1<u<0,
%\\
%w(-1)     &=& 1/\lambda,
%\end{array}
%\right.
\end{equation}
where the function $F\equiv F(u,w;\laa)$ is defined as:
\begin{equation}
\label{campoaux2}
F(u,w;\laa) = \dfrac{5}{2}u^{-1}+\gaa_{1}\left[ 1-\eta(-u)^{-\epsilon}  \right]+\gaa_{2}\left[ 1+\gaa_{3}\left(\laa+\dfrac{1}{u\,w}\right) \right]^{2} u\,.
\end{equation}

\noindent We have re-defined the eigenvalue of this problem as:
\begin{equation}\label{parametro}
\laa=\dfrac{c_{s}}{V_R} = \frac{1}{w(-1)}\, .
\end{equation}
Thus, the initial condition at the stellar surface for the EoM is,
\begin{equation}
\label{condini}
w(-1) = 1/\lambda \,.
\end{equation}

In order to find a transonic wind solution, i. e., a solution that start with a low speed value at the stellar surface and reaches a speed value larger than the sound speed at larges distances, the solution of this problem must pass through a singularity from the EoM. In addition, the dependency on $\lambda$ is nonlinear both in Eq.~\eqref{eom2} and also in the initial condition. Therefore, in the next section we will analyze the topology at the singularity.

\section{Topology of the EoM}
Following the detailed topological analysis of the CAK \citep{michel04b} and m-CAK \citep{michel07} cases, in order to obtain a transonic solution, the location of the singular point is obtained after the analysis of the singularity condition. In addition, to assure a smooth transition from the subsonic solution branch to the supersonic branch, an extra condition must be imposed, namely the regularity condition.

\subsection{Singularity and Regularity conditions}
Rearranging Eq.~\eqref{eom2} and defining a function $H$, we obtain:

\begin{equation}
\label{funcH}
H \equiv H(u,w,w';\laa) = \left(1-w^{2}\right)\,w' - w\,F(u,w;\laa)\, ,
\end{equation}
where $w'=dw/du$. Then, the mathematical definition of the singularity condition reads:
\begin{equation}
\label{singH}
\frac{\partial}{\partial w'}\,H = 0\,,
\end{equation}
what gives $(1-w^{2})=0$ or the corresponding decretion solution $w=+1$.

The regularity condition must be imposed at this singular (or sonic) point. This condition is:
\begin{equation}
\label{regH}
\frac{d}{du}\,H= \frac{\partial}{\partial u}\,H + w' \,\frac{\partial}{\partial w}\,H \, = 0.
\end{equation}
Equations \eqref{singH} and \eqref{regH} are only valid at the sonic point: $u=u_{s}$, $w(u_{s})=1$, and $w'(u_{s})=w'_s$. Solving $w'_s$ from Eq. \eqref{regH}, we obtain an analytical expression for the value of the velocity gradient ($w'_s$) at this point:
\begin{equation}
\label{wps}
w'_s=-\frac{1}{4}F_{w}(u_{s},1;\laa)\pm \dfrac{1}{4}\sqrt{  F_{w}(u_{s},1;\laa)^{2} -8 F_{u}(u_{s},1;\laa) }\,,
\end{equation}
here $F_{u}(u_{s},1;\laa)=\pp F/\pp u$ and $F_{w}(u_{s},1;\laa)=\pp F/\pp w$. Hereafter, we will use $F_u$ and $F_w$ instead of  $F_u(u_{s},1;\laa)$ and $F_{w}(u_{s},1;\laa)$, respectively. These partial derivatives of $F$ are the following:
\begin{eqnarray}
\label{Fu}
F_u &=&-\frac{5}{2 \, u^2} - \gamma_1 \,\eta \, \epsilon \,(-u)^{-(1+\epsilon)}\nonumber \\
   & & + \, \gamma_2 \left(1+\gamma_3 \left( \lambda+\frac{1}{u}\right)\right)^2 \nonumber \\
   & & - \, 2 \, \gamma_2 \,\gamma_3 \, u^{-1}
   \left(1 + \gamma_3 \left(\lambda
   +\frac{1}{u}\right)\right)
\end{eqnarray}
and 
\begin{equation}
\label{Fw}
F_w= -2 \,\gamma_2 \, \gamma_3 \,\left(1+ \gamma_3 \left(\lambda +\frac{1}{u}\right)\right).
\end{equation}

\subsection{Solution Branches}
A necessary condition for the existence of $w'_s >0$ (decretion) comes from argument of the square root in Eq.~\eqref{wps}, namely:
\begin{equation}
\label{restriccion2}
\Delta F = \left( F_{w}\right)^{2} -8\, F_{u} \geq 0.
\end{equation}

\noindent With this condition (Eq.~\ref{restriccion2}), from Eqs. \eqref{campoaux2} and \eqref{wps} there exist two branches for $w'_s$. 
Denoting these branches by $w'_{s+}$ and $w'_{s-}$,  where $w'_{s+} \geq w'_{s-}$. When $\Delta F\,=\,(F_{w})^{2} - 8 F_{u} =0$,  both are equal, $w'_{s+} = w'_{s-}$.
%%%%%%%%%%%%%%%%%%%%%%%%%%%%%%%%%%%%%%%%%%%%%%%%%%%%%
%%%%%%%%%%%%%%%%%%%%%%%%%%%%%%%%%%%%%%%%%%%%%%%%%%%%%
%%%%%%%%%%%%%%%%%%%%%%%%%%%%%%%%%%%%%%%%%%%%%%%%%%%%%
\subsection{Classification of the singular point \label{clasp}}
The  standard classification of singular points, following \cite{Amann90}, is shown in Table \ref{tipos}. To exemplify this classification we select, throughout this work,  the same stellar parameters from \cite{okazaki2001} for a B0 main-sequence star, i.e., $M = 17.8\,M_{\sun}$, $R = 7.41\,R_{\sun}$, and $T_{\rm eff} = 28\, 000\,K$.

\begin{table}[htp!]
\centering
\begin{tabular}{ccc}
	\hline\hline
Saddle &  & $F_u < 0$\\
Nodal &  & $0 \leq F_u \leq F_w^2 / 8$\\
Spiral &  & $F_u > F_w^2 / 8$\\
\hline
\end{tabular}
\caption{\small{Topology of the singular points.}
\label{tipos}}
\end{table}

\begin{figure}[hpt!]
\centering
\includegraphics[width=1.\linewidth]{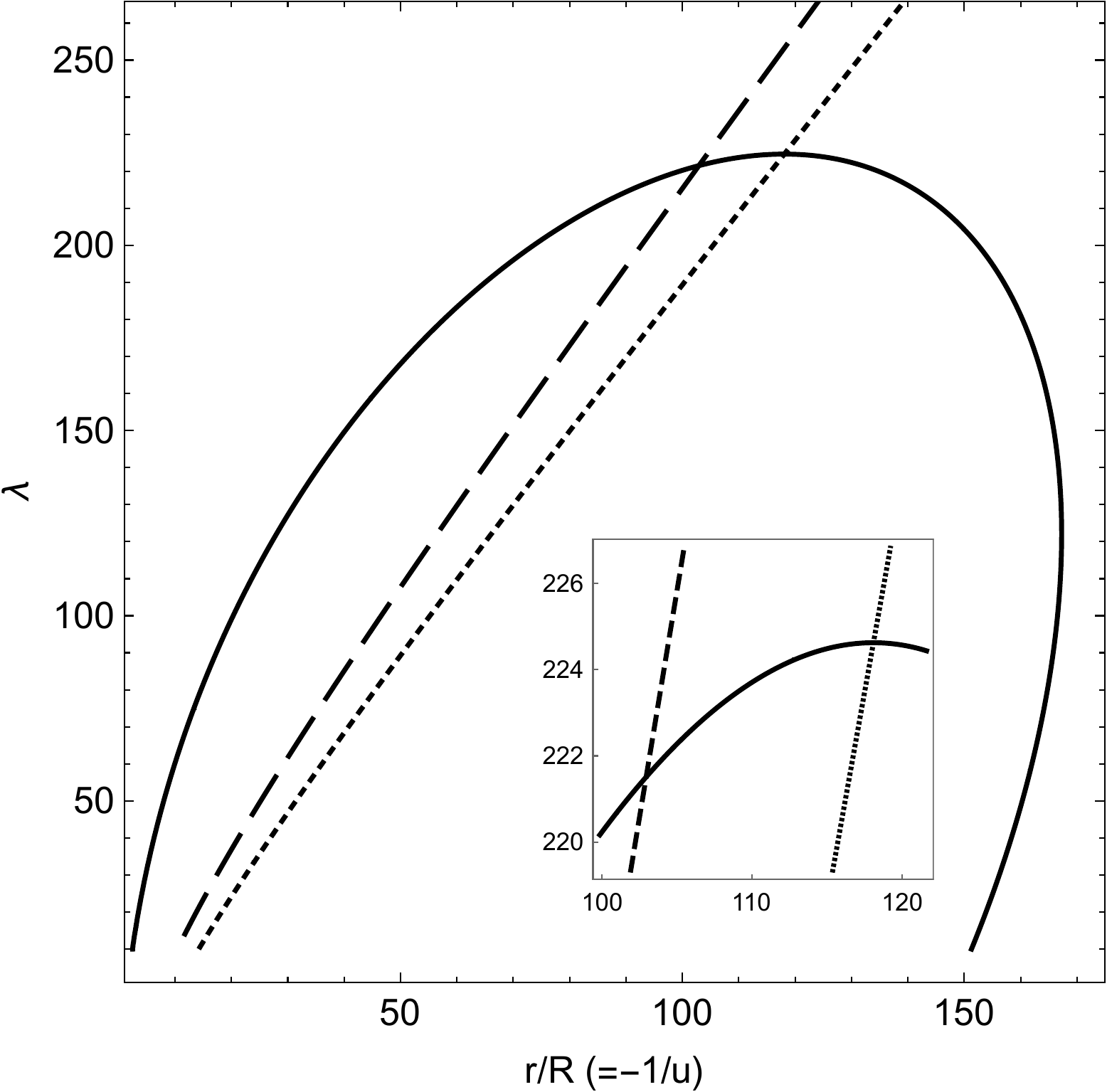}
\caption{\small{Different segments for the singular point classification, see text for details. We have used: $\Omega=0.9$, $\alpha=1.0$, $\epsilon = 0.1$, and $\eta = 0.5$.}}
\label{fig1}
\end{figure}

Figure \ref{fig1} shows contour plots in terms of $r/R$ and $\laa$ for the following conditions: black solid line corresponds to $F = 0$, $F_u = 0$ is shown in short-dashed straight line, and $F_{w}^{2} -8\, F_{u} = 0$ in long-dashed straight line. 
The Spiral topology corresponds to the segment of the curve $F=0$ that is at the left of the long-dashed straight line, Nodal topology corresponds to the segment that lies between both straight lines, and Saddle (or X-type) topology corresponds to the segment that is at the right of the short-dashed straight line. From the zoom of this figure, we see that we can restrict the range of the eigenvalue $\laa$ in order to solve numerically the EoM.

\subsection{Determining the range of the eigenvalues}
From Fig. \eqref{fig1}, defining as $\hat{r}_{s1}$ and $\laa_1$ as the values of $r_{s}/R$ and $\laa$ at the intersection between $F=0$ and $F_{w}^{2} -8\, F_{u} = 0$, we obtain $\hat{r}_{s1} = 65.74$ and $\laa_1 = 129.22$. Similarly for the other intersection, we obtain $\hat{r}_{s2} = 74.25$ 
and $\laa_2 = 130.82$. Using this information we can search for the values of the velocity gradient ($w'_{s\pm}$) and the eigenvalue ($\laa$) at the sonic point given by Eq. (\ref{wps}).
\begin{figure}[hpt!]
\centering
\includegraphics[width=1.0\linewidth]{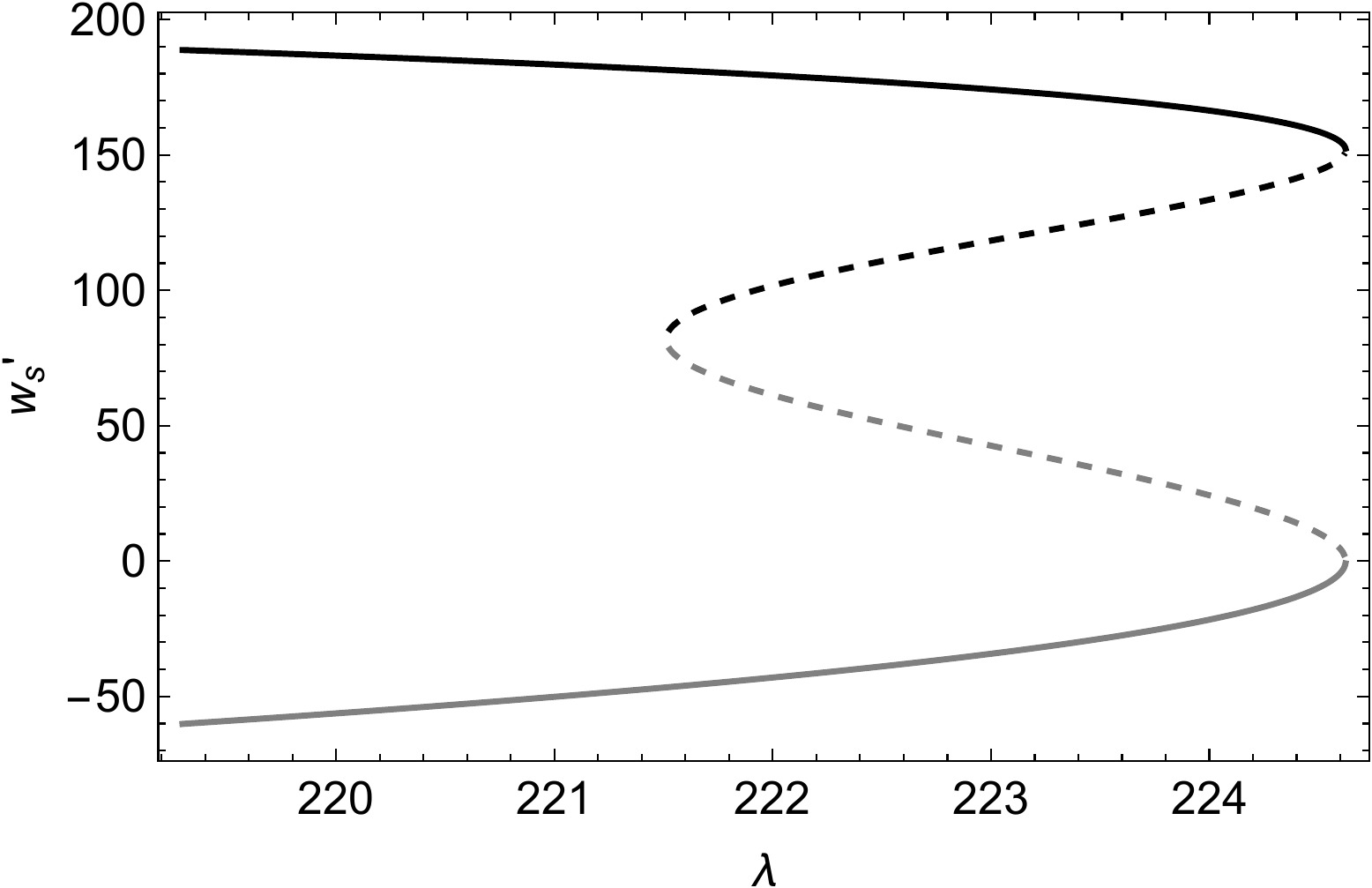}
\caption{\small{Velocity gradient at the singular point as function of the eigenvalue $\laa$.
Solid lines are for the Saddle solutions and dashed lines for the Nodal solutions. The parameter used are the same as Fig. \ref{fig1}. See text for details.}}
\label{fig2}
\end{figure}

Figure \ref{fig2} shows the value of $w'_{s}$ as function of the eigenvalue. 
In order to obtain this values we need to set the value of $\laa$ and then evaluate  
$w'_{s\pm}$ for values of $\hat{r}_s (=-1/u_s)$ in the range: $\hat{r}_{s1} < \hat{r}_{s} < \hat{r}_{s2}$ for the Nodal solution, shown in black dashed line for $w'_{s+}$ and in gray dashed line for $w'_{s-}$. On the other hand, for Saddle solution, we seek for  $\hat{r}_{s}$ in the range 
$ \hat{r}_{s} > \hat{r}_{s2}$. Solid black  line show the branch $w'_{s+}$ and gray solid line show the branch $w'_{s-}$, which is not a physical solution for a transonic decretion disk, because $w'_{s-}<0$.

\section{Numerical procedure}
Knowing the topology of the non-linear EoM, we can now describe the numerical procedure to obtain the different viscous transonic decretion disk solutions.
%\subsection{Nodal Solutions}

Our proposed procedure is the following:
\begin{itemize}
    \item[$\bullet$] Define a grid of eigenvalues ($\lambda_i)$ in the range ($\lambda_1$, $\lambda_2$) for Nodal solutions and ($a \lambda_1$, $\lambda_2$) for the Saddle solution. A typical value of $a$ is $0.7$, according our calculations.
    \item[$\bullet$] Integrate from the stellar surface $u=-1$, with the initial condition $w(-1)=1/\lambda_i$, up to $w(u)=1-\epsilon_0$. When this value of the velocity is attained, the location of the singular point, $u=u_s$ is therefore known. A typical value of  $\epsilon_0$ is  $10^{-6}$.
    \item[$\bullet$] For the entire grid of $\lambda_i$ solutions that reached the sonic point, we compare the values of the numerical velocity gradient at the sonic point, $(w'_{s})_i$, with the values of $w'_{s-}$ or $w'_{s+}$, depending on the solution type, obtained from Eq.(\ref{wps}), calculating the absolute error $\phi_i(\lambda_i)=\|(w'_{s})_i - w'_{s\pm}\|$.\\ It is worth to notice that when calculating $w'_{s\pm}$ from this Eq., the value of $u_s$ must lie in the segment  of the Nodal or Saddle solutions depending on the type of solution sought (see Fig. \ref{fig1}).
    \item[$\bullet$]  The eigenvalue $\lambda_i$ corresponds to $\min(\phi_i)$.
\end{itemize}

Base on this numerical methodology now we are in condition to perform numerical experiments to understand the behaviour of the solutions in terms of the different parameters involved in the EoM.

\section{Numerical calculations \label{calculos}}
In this section we will perform numerical calculations to understand the influence of the different parameters involved in the EoM. All our calculations hereafter are performed for a typical B0 main-sequence star: $M = 17.8\,M_{\sun}$, $R = 7.41\,R_{\sun}$, and $T_{\rm eff} = 28\, 000\,K$ with $T_{\rm disk}= 0.5 T_{\rm eff}$ and $\Omega=0.9$.
\subsection{Solutions without line force}
\begin{figure}[hpt!]
\centering
\includegraphics[width=1.0\linewidth]{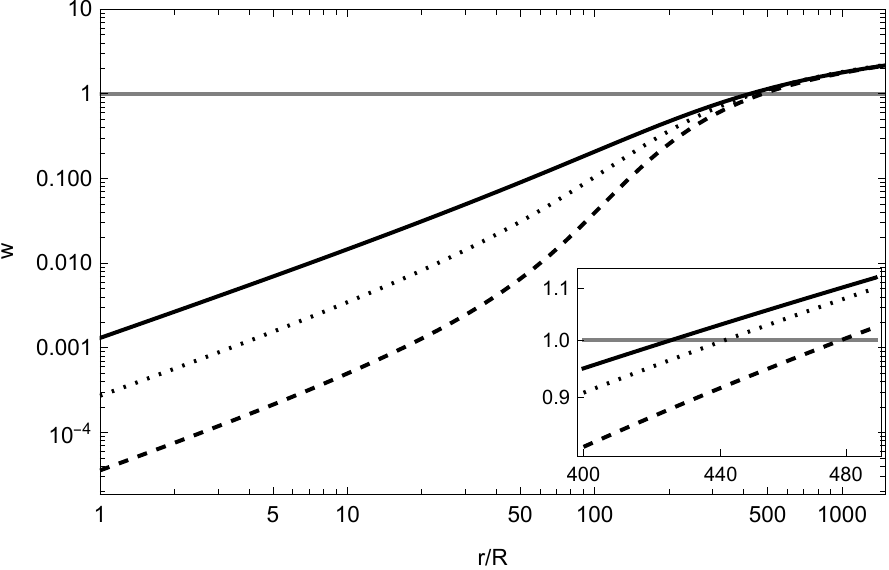}
\caption{\small{Dimensionless velocity profile, w, for the case without line force ($\eta=0$). Black solid line is for $\alpha =1.0$, dotted line for $\alpha =0.1$ and dashed line for $\alpha =0.01$. The zoom shows the region around the sonic point. Light gray horizontal solid line represents the sound speed value, $w=1$. Stellar parameters are given at the beginning of section \ref{calculos}. }}
\label{fig5}
\end{figure}

Following \cite{okazaki2001}, we solve first the EoM without line force, therefore we set $\eta=0$ to describe a \textit{pure} viscous outflow. Figure \ref{fig5} shows the normalized velocity profile, $w(r)$ as function of $r/R$. The black solid line correspond to $\alpha=1$, $\alpha=0.1$ in shown in black dotted line and $\alpha=0.01$ in black dashed line. Eigenvalues are $\lambda = 771.10; 3734.65; 28498.51$ for $\alpha=1.0;0.1;0.01$, respectively. The location of the sonic point is similar for all solutions and lie in the range $425 \leq r_s/R \leq 479$ (see the zoom in Fig. \ref{fig5}).
\subsection{The viscosity parameter $\alpha$}
Analogous to the results found by \cite{okazaki2001}, the main impact of the viscosity parameter $\alpha$ is the topology of the singular point. 
\subsubsection{Nodal solutions}
\begin{figure}[hpt!]
\centering
\includegraphics[width=1.0\linewidth]{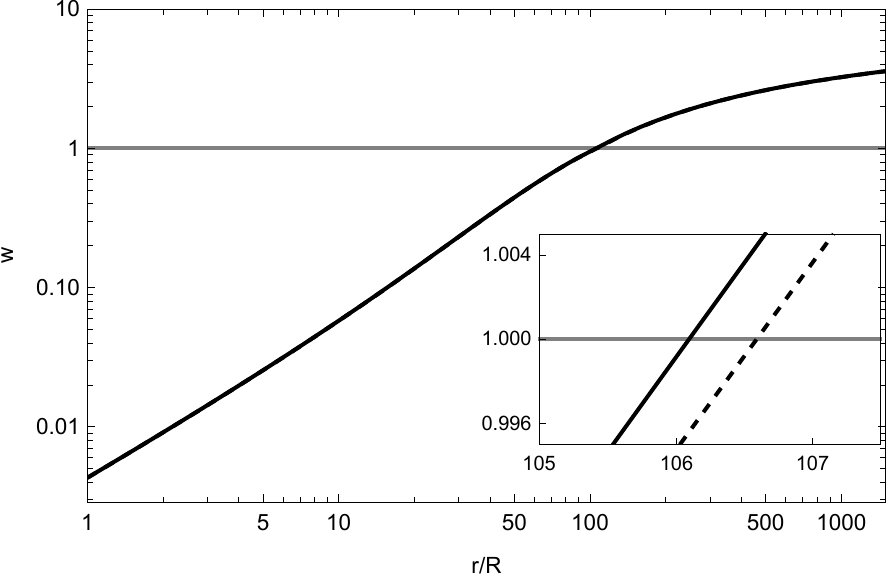}
\caption{\small{Dimensionless velocity profiles ($w$) for both  (indistinguishable) Nodal solutions as function of the $r/R$ coordinate, with $\alpha=1.0$ and line-force parameters $\epsilon=0.1$, $\eta=0.5$. The zoom shows both solutions around the singular point. Black solid line correspond to the $w'_{s+}$ branch solution and dashed line to the $w'_{s-}$ branch solution. Light gray solid line represents the sound speed value, $w=1$.Stellar parameters are given at the beginning of section \ref{calculos}}}
\label{fig3}
\end{figure}

High values of $\alpha$ implies that the sonic point has a Nodal topology. This topology shows two branches, shown in dashed lines in Fig. \ref{fig2}.
Figure \ref{fig3} show the velocity profiles for both Nodal solutions. Viscosity and line force parameters are: $\alpha=1.0$, $\epsilon=0.1$, $\eta=0.5$, and $T_{\rm disk}=  0.5 \, T_{\rm eff}$.
Both Nodal solutions are indistinguishable from each other on the scale of Fig. \ref{fig3}, but in the zoom around the singular point, it is possible to distinguish them separately.
The solution with $w'_{s+}$, in black solid line, has $\lambda= 223.15$ and $r_s =107.78\, R$; the  $w'_{s-}$ solution, dashed line,  has $\lambda= 222.91$ and $r_s = 106.93\, R$.
For any practical issue, there are almost no difference in the behaviour of both Nodal solutions and both have almost the same eigenvalue, therefore we can select any of them.

\subsubsection{Saddle solution}
Depending on the value of $\alpha$, the topology switch between Saddle and Nodal.% (see Fig.\ref{figetaalpha} below \textbf{será bueno adelantarse con el orden de las figuras?}).

\begin{figure}[hpt!]
\centering
\includegraphics[width=1.0\linewidth]{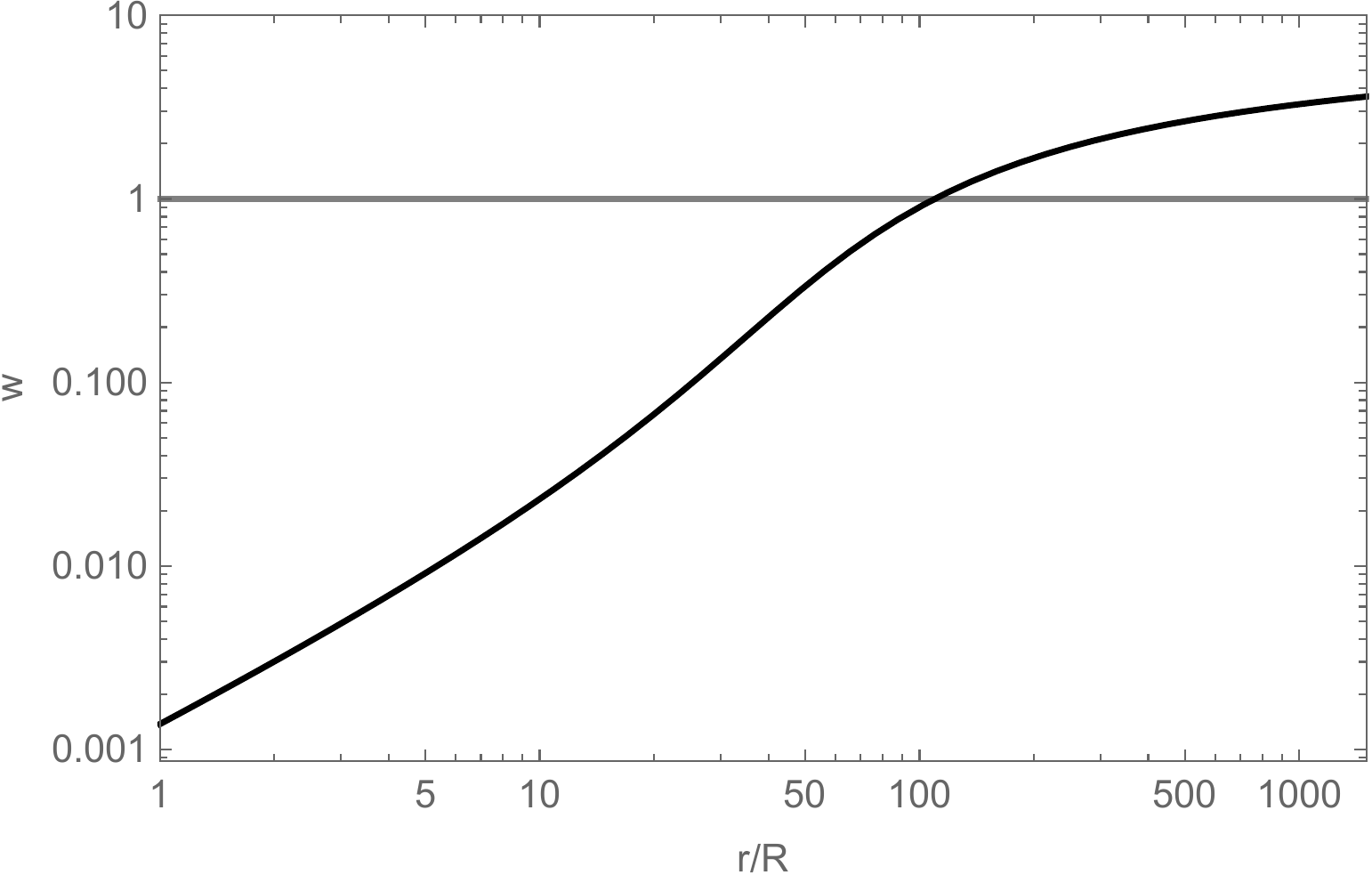}
\caption{\small{Dimensionless velocity profile ($w$) in solid black line for the Saddle solution with $\alpha=0.2$ and line-force parameters $\epsilon=0.1$, $\eta=0.5$. Light gray solid line represents the sound speed value, $w=1$. Stellar parameters are given at the beginning of section \ref{calculos}. }}
\label{fig4}
\end{figure}

Using same parameters as the previous Nodal case, but with $\alpha=0.2$, the resulting Saddle solution is shown in Fig. \ref{fig4}. We clearly see here that the location of the sonic point is almost the same as both Nodal solutions, here we have $\lambda= 679.63$ and $r_s = 109.62 \, R$.

Although the Nodal and Saddle solutions, shown in Fig. \ref{fig3} and Fig. \ref{fig4}, only differ in the values of the viscosity parameter $\alpha$ and, consequently, in their topology, the location of the sonic point hardly differs and the velocity profiles $w(r)$ are very similar.
%This is a different result of the one obtained from \cite{okazaki2001}}

In the next sub-sections we investigate in detail the role of the different parameters in this viscous transonic decretion outflows.

\subsection{Type of solution depending on the value of $\eta$ \label{depEta}}
\begin{figure}[hpt!]
\centering
\includegraphics[width=0.9\linewidth]{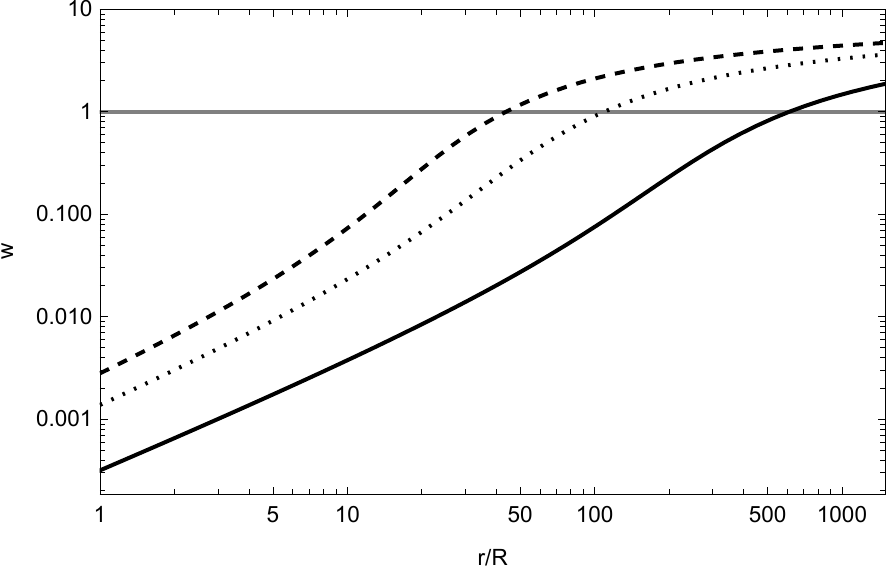}
\caption{\small{Dimensionless velocity profile, $w$ as function of $r/R$ with $\epsilon=0.1$ and $\alpha=0.2$. Here $\eta=0.1$ is shown in black solid line, $\eta=0.5$ in black dotted line, and $\eta=0.6$ in  black dashed line. Light gray solid line represents the sound speed value, $w=1$. Stellar parameters are given at the beginning of section \ref{calculos}}}
\label{fig6}
\end{figure}

The line force strength and behavior is determined by the $\eta$ and $\epsilon$ parameters, respectively. Here we study the variation of the $\eta$ parameter with $\alpha=0.2$. Figure \ref{fig6} shows $w(r)$ for different values of the $\eta$ parameter with $\epsilon=0.1$. We confirm that the larger is $\eta$, the nearer (from stellar surface) is located the sonic point. The specific locations are: $r_s/R=614.8; 109.6; 43.9$ for $\eta=0.1; 0.5; 0.6$, respectively. When $\eta=0.6$ the sonic point is located at a distance lower than $50\,R$, but such a strong force is very unlikely. Below (section~\ref{sect55}) we discuss other combinations of line-force parameters and disk temperatures that give similar results about the location of the singular point.

\begin{figure}[hpt!]
\centering
\includegraphics[width=1.0\linewidth]{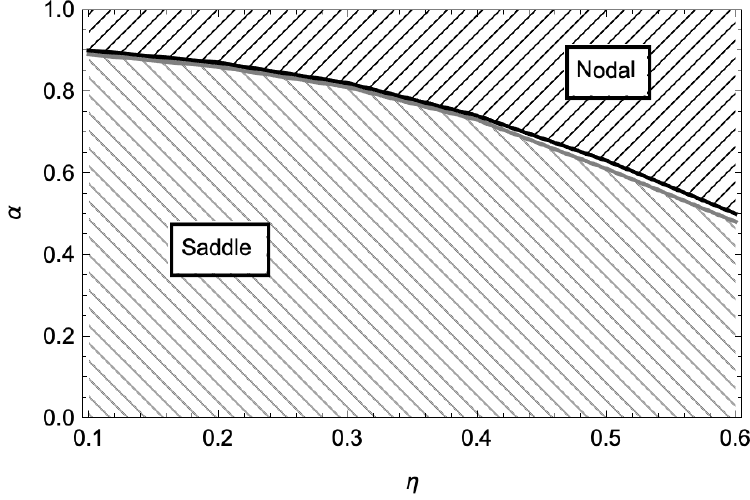}
\caption{\small{$\alpha$ values for Nodal and Saddle regions as function of $\eta$ with $\epsilon=0.1$. Solid black continuous line denotes the lower limit of $\alpha$ for Nodal solutions. Light gray continuous line denotes the upper limit of Saddle solutions. Hatched zones represents the Nodal (upper) and Saddle (lower) regions. Stellar parameters are given at the beginning of section \ref{calculos}. See text for details.}}
\label{figetaalpha}
\end{figure}

 We studied in terms of $\alpha$ parameter when the solution is Nodal or Saddle as function of the $\eta$ line force parameter. We find that the switching zone is very narrow as it is shown in Fig. \ref{figetaalpha} and summarized in Table\ref{tabla2}.\\ %\textbf{que sucede en la region intermedia???} 

 All the solutions we obtained from the  EoM are physical solutions. The characterization of the stability of the steady state can be analyzed by means of the evolution of perturbative waves in the time-dependent equation of motion \citep[see different approaches in][and references therein]{criminale2018}, this type of study is beyond the scope of this work.
\begin{table}[htp!]
\centering
\begin{tabular}{c|c|c}
	\hline\hline
$\eta$ & \multicolumn{2}{c}{$\alpha$}   \\
\cline{2-3}
       & Nodal &  Saddle\\
\hline
0.6 & $>\,0.50$ & $<\,0.48$\\
0.5	& $>\,0.63$ & $<\,0.61$\\
0.4	& $>\,0.74$	& $<\,0.73$\\
0.3	& $>\,0.82$	& $<\,0.81$\\
0.2	& $>\,0.87$	& $<\,0.86$\\
0.1	& $>\,0.90$	& $<\,0.89$\\
\hline\hline
\end{tabular}
\caption{\small{Topology of the singular points in terms of $\alpha$ and $\eta$. Stellar parameters are: $M = 17.8\,M_{\sun}$, $R = 7.41\,R_{\sun}$, and $T_{\rm eff} = 28\, 000\,K$ with $T_{\rm disk}= 0.5 T_{\rm eff}$ and $\Omega=0.9$.}
\label{tabla2}}
\end{table}

\subsection{Type of solution depending on the value of $\Omega$}
\begin{figure}[hpt!]
\centering
\includegraphics[width=1.0\linewidth]{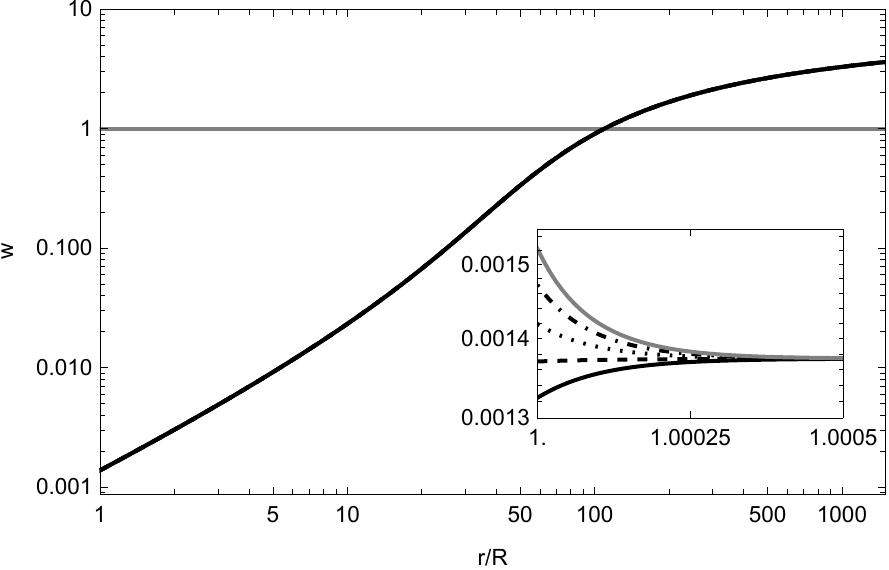}
\caption{\small{Dimensionless velocity profile, w as function of $r/R$ for different values of the $\Omega$ with $\alpha=0.2$, $\epsilon=0.1$
and $\eta=0.5$. In gray solid line $\Omega=0.99$ is shown, $\Omega=0.9$ in dot-dashed line,
$\Omega=0.8$ in dotted line, $\Omega=0.7$ in dashed line, and $\Omega=0.6$ in black solid line. Light horizontal gray solid line represents the sound speed value, $w=1$. Stellar parameters are given at the beginning of section \ref{calculos}. See text for details.}}
\label{fig7}
\end{figure}

\begin{figure}[hpt!]
\centering
\includegraphics[width=1.0\linewidth]{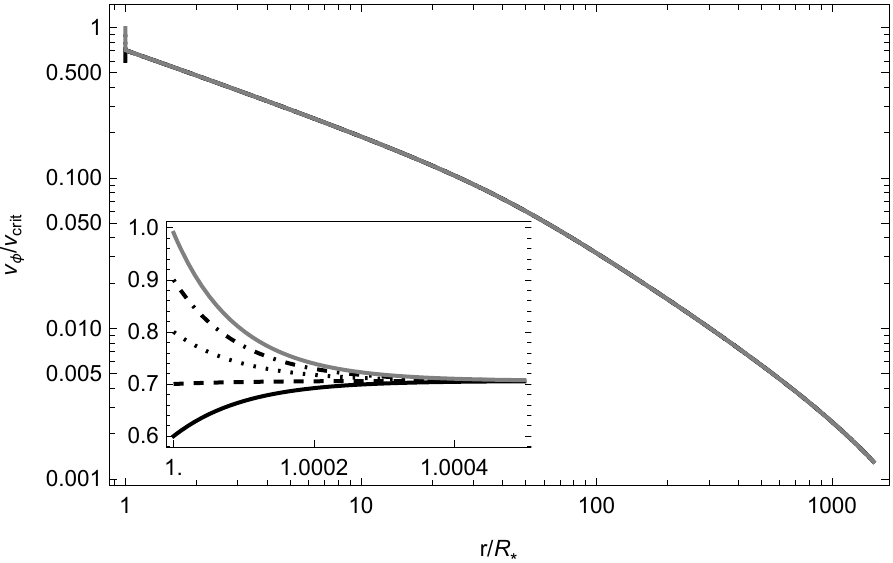}
\caption{\small{$V_{\phi}(r)/V_{\rm crit}$ as function of $r$ for the same set of parameters as Fig.  \ref{fig7}. All solutions converge to one in a very small region above the stellar surface. See text for details.}}
\label{fig8}
\end{figure}

In order to study the influence of the centrifugal force on this viscous disk decretion model, we calculate for different values of $\Omega$ the dimensionless velocity profile $w$ as function of $r$ (see Fig.~\ref{fig7}). These $w(r)$ profiles show a very unexpected result, all of them converge to an unique solution after a very small distance above the stellar surface, $r < 1.0004 R$ as shown in the zoom of this Figure. Similarly, Fig.~\ref{fig8} shows the behavior of $V_{\phi}(r)/V_{\rm crit}$ as function of $r$. Again, all solutions converge to an unique solution very near the stellar surface. 

A steep gradient in $V_{\phi}(r)$ might cause a boundary layer, provoking that the specific angular momentum just outside of this layer is settled into a value, which is smoothly connected to the outer parts of the disk. %This means that the specific angular momentum (or the rotational rate $\Omega$) at the inner edge of the disk (stellar surface in this model) is not longer a free parameter. 
The results shown in Fig.~\ref{fig8} can be used to constraint the value of $\Omega$ (or the specific angular momentum) to a much smaller region, i.e., obtaining a smooth solution, in this case $\Omega \sim 0.7$, where the gradients are not steep. 

\subsection{Type of solution depending on the line-force parameters and $T_{\rm disk}$ \label{sect55}}
We analyze the behaviour of the $\epsilon$ parameter and the disk temperature, $T_{\rm disk}$, in the VDD model. The parameter  $\epsilon$  describe the decay of the line force as we move outwards from the stellar surface.
\begin{table}[htp!]
\centering
\begin{tabular}{c|c|c}
	\hline\hline
$\epsilon$ & $r_{s}/R$ & $\lambda$  \\
\hline
0.1 & 614.77 & 3179.91\\
0.3	& 272.34 & 1864.86\\
0.5	& 45.45	& 716.68\\
\hline\hline
\end{tabular}
\caption{\small{Location of the sonic point and its corresponding eigenvalue, $\lambda$, in terms of the line-force parameter $\epsilon$ with $\alpha=0.2$ and $\eta=0.1$. Stellar parameters are given in Table \ref{tabla2}.}
\label{tabla3}}
\end{table}

Table \ref{tabla3} summarises the location of the sonic point and eigenvalue in terms of the value of $\epsilon$. The dependence of the line force in terms of $r$, when $\Gamma \to 0$, is given by (see Eq. \ref{cheneq}):
\begin{equation}
g_{\rm rad} \to \frac{G M}{r^2} \, \eta \, \left(\frac{r}{R}\right)^\epsilon.
\label{cheneq2}
\end{equation}
Thus, the larger is $\epsilon$ the slower is its decay in terms of $r$ and due to a larger line-force, the location of the sonic point lies closer to the stellar surface.\\

Table \ref{tabla4} summarises the location of the sonic point and Eigenvalue in terms of the value of the disk temperature, $T_{\rm disk}$  and different combinations of  line-force parameters  $\eta$ and $\epsilon$. %Here, we have use $\epsilon=0.5$  and $\eta=0.1$, for the models of Table \ref{tabla4}. 
The influence of $T_{\rm disk}$ in the location of the sonic point is not determinant when the value of $\epsilon$ is high. On the other hand, for low values of $\epsilon$, the influence of $T_{\rm disk}$ is quite important, reducing the location of $r_s$ from $\sim240\,R$ to $\sim160\,R$ when $T_{\rm disk}/ T_{\rm eff}$ increases from 0.5
to 0.9, when $\eta = \epsilon =0.2$.

\begin{table}[htp!]
\centering
\begin{tabular}{c|c|c|c|c}
	\hline\hline
 $\eta$& $\epsilon$ & $T_{\rm disk} / T_{\rm eff} $& $r_{s}/R$ & Eigenvalue   \\
\hline
0.1 & 0.5 & 0.5 & 45.45 & 716.68\\   
0.1 & 0.5 & 0.7	& 44.18 & 591.76\\
0.1 & 0.5 & 0.9	& 42.83	& 509.99\\
0.2 & 0.2 & 0.5 & 243.99 & 1576.36\\
0.2 & 0.2 & 0.7 & 192.11 & 1181.94\\
0.2 & 0.2 & 0.9 & 159.68 & 947.98\\
0.3 & 0.2 & 0.5 & 101.98 & 833.21\\
0.4 & 0.2 & 0.5 & 36.53 & 383.21\\
0.5 & 0.2 & 0.5 & 13.25 & 139.58\\
0.5 & 0.2 & 0.7 & 13.03 & 115.17\\
0.5 & 0.15 & 0.5 & 35.17 & 308.18\\

\hline\hline
\end{tabular}
\caption{\small{Location of the sonic point and its corresponding eigenvalue ($\lambda$) in terms of the disk temperature $T_{\rm disk}$ and line-force parameters  $\epsilon$, $\eta$. Stellar parameters are $M = 17.8\,M_{\sun}$, $R = 7.41\,R_{\sun}$,  $T_{\rm eff} = 28\, 000\,K$ and $\Omega=0.9$. }
\label{tabla4}}
\end{table}

%\subsection{Other line-force parameters combinations From the results shown in Table \ref{tabla4},}
As we pointed out  in section~\ref{depEta}, it is very unlikely to have such a strong line-force with $\eta=0.6$. However, it is indeed possible with plausible values of $\epsilon$, $\eta$ and  $T_{\rm disk}$ to obtain transonic disk solutions that might explain the results of \cite{Klement17}.

\section{Discussion}
\begin{figure}[hpt!]
\centering
\includegraphics[width=1.0\linewidth]{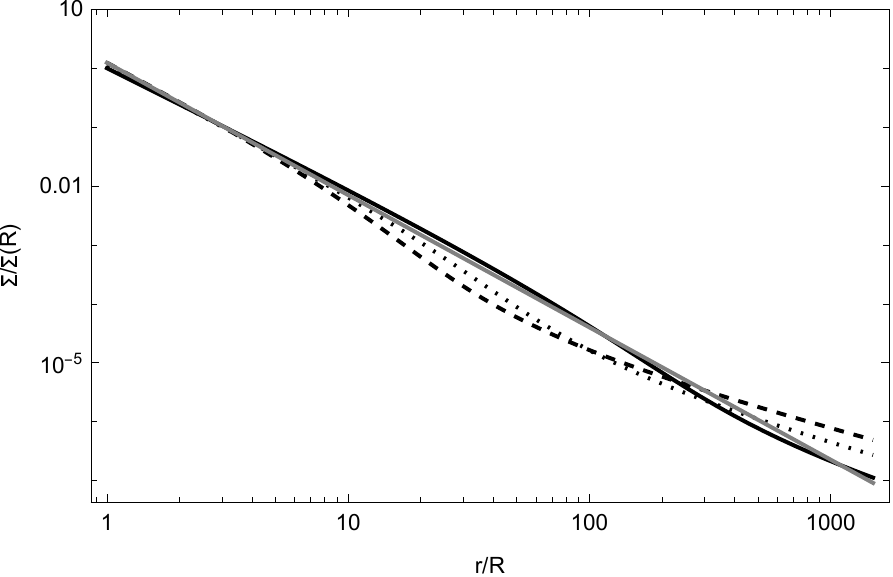}
\caption{\small{$\Sigma(r)/\Sigma(R)$ as function of $r$ with $\alpha=0.2$ and $\epsilon=0.1$, all stellar parameters are given at the beginning of section \ref{calculos}. Here $\eta=0.1$ is shown in black solid line, $\eta=0.5$ in black dotted line and $\eta=0.6$ black dashed line. Light gray solid line represents the fit of Eq. \ref{fitSigma} for the $\eta=0.6$ case. See text for details.}}
\label{fig9}
\end{figure}

In this work we have shown the dependence of the velocity field for different parameters involved in the EoM from the VDD model. However, the standard methodology used to obtain an observable, such as H$\alpha$, is to use the (volumetric) density ($\rho$) or  the vertical integrated density ($\Sigma$), as function of $r/R$ in a Keplerian orbit as input in radiative transport codes, such as, BEDISK \citep{Sigut2007} or HDUST \citep[][]{Carciofi2006}.
The standard modelling of $\rho$ is:
\begin{equation}
    \rho(r)=\rho_0 \left(\frac{r}{R}\right)^{-m} e^{-(z/H)^2} \, ,
\end{equation}
where $\rho(R)=\rho_0$ is the density at the stellar surface.
The definition for $\Sigma$, follows after integrate $\rho(r)$ in the $z$ direction, namely:
\begin{equation}
    \Sigma(r)=\Sigma_0 \left(\frac{r}{R}\right)^{-n}\, ,
\label{fitSigma}
\end{equation}
here\footnote{for isothermal disks, $H(r) \propto r^{3/2}$ } $n=m-3/2$ and $\Sigma(R)=\Sigma_0$, is the vertical integrated density at the base of the wind.\\
Figure \ref{fig9} shows different $\Sigma(r)/\Sigma(R)$ as function of $r$ for the models shown in section \ref{depEta} (see also Fig. \ref{fig6}).  The solid gray line represent the fit of Eq. \ref{fitSigma} to the vertical integrated density profile for $\eta=0.6$. We clearly see in this figure, that none of the solutions maintains the behaviour described by Eq. \ref{fitSigma}  in the entire range of $r$, but only in the range $R<r \lesssim 10 \,R$. The results for the fits are:
$n=0.745$ for $\eta=0.6$, $n=0.649$ for $\eta=0.5$, and $n=0.554$ for $\eta=0.1$.
All these values of $n$ are calculated in the interval $R\leq r \leq 20 \,R$. Fits for the interval $R\leq r \leq 100 \,R$, give values of $n$ that differs in the fourth decimal with the previous fit interval. These results shows that the standard formulae giving by $\rho(r)$ or $\Sigma(r)$ are fairly good approximations to our results, but only for $r\lesssim 10\, R$.\\

A possible explanation about the different behaviours  % of $\Sigma(r)/\Sigma(R)$ %
shown in Fig. \ref{fig9}, with different locations of the sonic point for each $\eta$ (see section \ref{depEta}) is the following: even if the Shakura-Sunyaev viscosity model is applicable to supersonic regions, the viscosity is just inefficient there as angular momentum transfer mechanism, because the advection timescale in the supersonic region is much shorter than the viscous timescale, which makes the flow to be angular-momentum conserving.\\

In addition, \cite{Klement17} studied the observed SED turn-down by means of the VDD model, concluding that it can be explained with a truncated disk. They argued about two possible explanations of this: tidal forces from a close binary companion or a velocity profile with a transonic transition not too far from the star. They assumed that a binary companion is the most probable scenario, but for most of their six sources  binarity remained undetected. Nevertheless, we showed in this work, that for a suitable combination of $\eta$, $\epsilon$ and/or $T_{\rm disk}$, it is possible 
to have singular point locations that explain the SED turn-down without needing a close companion.

In reference on the influence of the stellar rotational speed (in terms of the critical speed), $\Omega$, we found that viscosity effects collapse,  in a very small region above the stellar surface, all the solutions to an (almost) unique $w(r)$ and $v_{\phi}(r)$ profiles as shown in Fig. \ref{fig7} and Fig. \ref{fig8}.The results from $v_{\phi}(r)$ shown in Fig. \ref{fig8}, can be used to restrict the range of $\Omega$  leading to a smooth solution, in order to avoid the formation of a boundary layer.

\section{Conclusions}
We have revisited \cite{okazaki2001} work of a Viscous Transonic Decretion Disk model with the inclusion of a radiative acceleration produced by an ensemble of optically thin lines, described by the  \cite{chen1994} model.
We developed a new solution approach, where the Eigenvalue of the problem is not longer an angular quantity but a radial one. After a detailed topological analysis of the steady-state Equation of Motion, three possible physical solution were found: 2 Nodal solutions, where the viscosity parameter $\alpha$ is larger (Nodal region in Fig.~\ref{figetaalpha}) and one Saddle solution for lower values of $\alpha$ (Saddle region in Fig.~\ref{figetaalpha}). Both Nodal solutions are almost indistinguishable between them.
The value of the viscosity parameter $\alpha$, given by the Shakura–Sunyaev model, is not determinant for the location of the sonic point. % {\bf{and shape of the velocity profile $w(r)$}}. 
Other parameters, especially those of the line force, $\eta$ and $\epsilon$, directly influence the solution of the EoM.
Finally, any effect of the stellar rotation is rapidly damped close to the stellar surface due to viscosity. In addition,  in order to obtain only smooth solutions, the range of $\Omega$ should be restricted.

As a future work, we will describe the line force using the standard CAK (and its improvements) theory \citep{cak} instead of the {\it ad-hoc} model from \cite{chen1994}. With this more detailed model for the line acceleration, it will be possible to incorporate rapid rotational effects such as oblate shape and gravity darkening to better describe a transonic VDD model of Be stars.

Finally, to study the observable such as the SED or the line H$\alpha$ \citep[see][]{Klement17} we plan to use the density profile, $\Sigma(r)$, as input in BEDISK and/or HDUST to calculate these observable.

	%_____AGRADECIMIENTOS___________________________________________________________________________
\begin{acknowledgements}
The authors would like to thank the referee, Atsuo Okazaki, for
his thoughtful comments and suggestions to improve this work. MC \& CA acknowledge the support from Centro de Astrof\'isica de Valpara\'iso.
MC, CA \& IA thanks the support from FONDECYT project 1190485.  MC and CA
thank to project ANID-FAPESP 2019/13354-1. IA is also grateful for the support from FONDECYT project 11190147. CA thanks the support from FONDECYT project 11190945.
This project has also received funding from the European Unions Framework Programme for Research and Innovation Horizon 2020 (2014-2020) under the Marie Sk{\l}odowska-Curie grant Agreement N$^{\rm o}$ 823734.
This work has been possible thanks to the use of AWS-U.Chile-NLHPC credits. Powered@NLHPC: This research was partially supported by the supercomputing infrastructure of the NLHPC (ECM-02). 
\end{acknowledgements}
	
	%_____BIBLIOGRAFÍA_______________________________________________________________________________
	\bibliography{biblio.bib} % your references Yourfile.bib
	\bibliographystyle{aa} % style aa.bst
	%\end{multicols}
	
	\begin{appendix} %First appendix
		
	\end{appendix}
\end{document}